\documentclass[pra,twocolumn,a4paper,showpacs,superscriptaddress]{revtex4}

 \usepackage{latexsym}
 \usepackage{amsmath}
 \usepackage{amsfonts}
 \usepackage{graphicx}
 \usepackage{amssymb}
 \usepackage{subfigure}
 \usepackage{color}

 \newcommand{\ket}[1]{\ensuremath{|#1\rangle}}
 \newcommand{\bra}[1]{\ensuremath{\langle #1 |}}

 \newcommand{\bc}{\begin{center}}
 \newcommand{\ec}{\end{center}}
 \newcommand{\mf}[1]{\boldsymbol{#1}}
 
 \newcommand{\ii}{i}
 \newcommand{\DP}{\Delta_1}
 \newcommand{\DC}{\Delta_2}
 \newcommand{\DM}{\Delta_3}

\begin{document}

\title{Enhancement of image resolution beyond the diffraction-limit by interacting dark resonances}

\author{Onkar N. \surname{Verma}}
\email{onkar@iitg.ernet.in}
\affiliation{Department of Physics, Indian Institute of Technology Guwahati, Guwahati- 781 039, Assam, India}
\author{Tarak N. \surname{Dey}}
\email{tarak.dey@gmail.com}
\affiliation{Department of Physics, Indian Institute of Technology Guwahati, Guwahati- 781 039, Assam, India}
\date{\today}

\begin{abstract}
We show how quantum coherence effects can be used to improve the resolution and the contrast of diffraction-limited images imprinted onto a probe field. 
The narrow and sharp spectral features generated by double-dark resonances (DDR) are exploited to control absorption, dispersion and diffraction properties of the medium.
The spatial modulated control field can produce inhomogeneous susceptibility of the medium that encodes the spatial feature of the control image to probe field in the presence of DDR.
The transmission of cloned image can be enhanced by use of incoherent pump field.
We find that the feature size of cloned image is four times smaller than the initial characteristic size of the control image even though the control image is completely distorted after propagation through 3 cm long Rb vapour cell. 
We further discuss how spatial optical switching is possible by using  of induced transparency and absorption of the medium.
\end{abstract}
\maketitle
\section{Introduction}
The ability to enhance spatial resolution of a Rayleigh or Sparrow limited image is one of the main challenges in optics \cite{Rayleigh}.
Conventional optics has failed to resolve the characteristic size of an image beyond a value comparable to the wavelength of the probing light \cite{saleh}. 
Main constraint of high resolution imaging comes from the diffraction and the absorption.
The diffraction of an image is inevitable due to its geometrical origin \cite{born}.
The above obstacles can completely or partially be eliminated  by use of quantum interference effects. 

Coherent electromagnetic fields interacting in a multilevel atomic system induce atomic coherence.
The induced atomic coherence can be exploited to demonstrate many interesting phenomena such as coherent population trapping (CPT) \cite{Arimond_po_96}, electromagnetically induced transparency (EIT) \cite{harris,michael}, lasing without inversion (LWI) \cite{olga_92} and saturated absorption techniques \cite{hansch_71,agarwal_lpr_09}.
A suitable spatially-dependent profile of the control field can produce a waveguide-like structure inside the medium which controls image propagation without diffraction\cite{truscott_99,kapoor_00,howell_09,jorg_11}.
This spatially varying refractive index can also guide focusing\cite{Moseley1,Moseley2,focusing3,focusing4,mitsunaga}, de-focusing\cite{defocusing}, self imaging\cite{cheng_07} and steering of the probe beam \cite{lida}. 
Most of the schemes employ a spatially inhomogeneous control field to protect the image from diffraction.
In a different development, Firstenberg {\it et al.} theoretically and experimentally found that Dicke narrowing induced by atomic motion and velocity-changing collisions is useful to eliminate the diffraction of an arbitrary image \cite{Firstenberg1,Firstenberg2,Firstenberg3}. 

Tailoring the optical properties of the medium along the transverse direction can open up a new possibilities of transferring the characteristics features of the control field to the probe field.
This is because the propagation dynamics of probe field is dependent on the diffraction and dispersive properties of the medium.  
The diffraction and dispersion characteristics of the atomic medium can be manipulated by using proper spatially inhomogeneous control field.
This concept has been demonstrated in both experimentally \cite{Li} and theoretically \cite{om} in a CPT system where well resolved control field structure used for optical cloning.
Further, the transmitted cloned image has feature size four times smaller as compared to the initial control image. 
However, all of these schemes suffer from strong absorption due to breaking of two-photon resonance condition. 
Hence the absorption based mechanism limits practical implementation. 
Therefore, one can take advantages of gain based schemes to generate high resolution cloned image.
Resolution of cloned image  can be improved by engineering the contrast of the refractive index of atomic waveguides of the gain medium.
Quantum interference effects induced by interacting dark resonances have  been shown to drastically increase the contrast of the refractive index profile \cite{chris,Lukin}.

In this paper, we have used interacting dark resonances to imprint the Rayleigh limited or Sparrow limited control image to probe field with high resolution and contrast. 
To facilitate these processes, we use four-level atomic system.
A single dark state can be created by  the control and the probe fields couple to the two arms of $\Lambda$-system.
This interaction gives rise usual single transparency window.  
The double-dark states can generate by using a microwave or optical field which interacts with a magnetic or electric dipole moments of relevant atomic transitions \cite{chen,yelin,ye}.
We find that the interference between two dark states results a new sharp absorption peak at line centre. 
The double dark resonance(DDR) spectra shows two transparency windows accompanied with one sharp absorption peak. 
Furthermore, we demonstrated that a very weak incoherent pump field is sufficient to turn the induced absorption dips to gain peaks. 
We exploit these sharp spectral features to write waveguide inside medium. 
We begin with Rayleigh limited control field structure and do a comparative study of inhomogeneous susceptibility for EIT, Microwave induced absorption (MIA), and LWI. 
The result shows that the presence of three fields with an incoherent pump provides a sharp contrast in refractive index from core to cladding than other two cases. 
We efficiently use  this sharp refractive index contrast  for cloning the Rayleigh limited control field image to the probe field with high resolution. 
Finally, we also show that Sparrow limited three modes of the control image can also be cast onto the probe field with appreciable resolution and high transmission. 
Later, we also use induced absorption and transparency mechanism to demonstrate the spatial switching (off or on) of probe beam. 
The spatial optical beam switching based on spatial phase modulation has been discussed recently in optical lattice \cite{hong}. 

The organisation of the paper is as follows.
In the next section, we introduce our model configuration, discuss the equations of motion for four-level system, 
describe the perturbative analysis of linear susceptibility of the probe field and derive the beam propagation equations for both probe and control fields under paraxial approximations. 
In Sec. III, we present our results. 
First, we describe the linear response of the medium to the probe field under the action of the continuous wave(cw) as well as the spatially dependent control beam.
We then employ the spatial dependent susceptibility to explain the basic principle of  cloning of Rayleigh limited control image to the probe field with high resolution and high contrast.
Next we provide numerical results on propagation dynamics of cloned images with different spatial structure of the control field for LWI, EIT and MIA cases.
Sec. IV provides a summary and discussion of our results.
\section{Theoretical Formulations}
\subsection{Model configuration}
\begin{figure}[t]
\includegraphics[width=0.8\columnwidth]{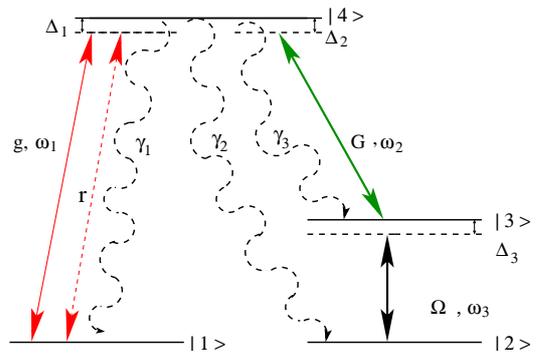}
\caption{\label{fig:Fig1} (Color online) 
 Schematic diagram of the four-level $^{87}$Rb atomic system.  The atomic transition $\ket{4}\leftrightarrow\ket{1}$ is coupled by the coherent probe field $g$ and incoherent pump field $r$. The control field $G$ interacts to the atomic transition $\ket{4}\leftrightarrow\ket{3}$. A microwave field $\Omega$ acts on the transition $\ket{3}\leftrightarrow\ket{2}$ to produce the double dark resonance of the system.}
\end{figure}

In this work, we consider a homogeneously broadened four level atomic system consisting of an excited state $\ket{4}$ and three metastable states $\ket{1}$, $\ket{2}$, and $\ket{3}$ interacting with two optical fields and one microwave field as shown in the Fig.~\ref{fig:Fig1}. 
The excited state $\ket{4}$  is coupled to two degenerate ground states $\ket{1}$, and $\ket{3}$ by two coherent fields, namely, a weak probe field with frequency $\omega_1$ and a control field with frequency $\omega_2$, respectively, which form a three level $\Lambda$-system. 
 The ground state $\ket{3}$ is further coupled to the metastable state $\ket{2}$ by an additional microwave field with frequency $\omega_3$. 
 We define two co-propagating optical fields along the $z$-axis as
\begin{equation}
\label{field}
 {\vec{E}_j}(\vec{r},t)= \hat{e}_{j}\mathcal{E}_{j}(\vec{r})~e^{- i\left(\omega_j t-  k_j z\right )} + {c.c.}\,,
\end{equation}
where, $\mathcal{E}_{j}(\vec{r})$ is the slowing varying envelope, $\hat{e}_{j}$ is the unit polarization vector, $\omega_j$ is the laser field frequency and $k_j$ is the wave number  of field, respectively.  The index $j\in \{1,2\}$ denotes the probe or control field, respectively. The microwave field is defined as 
\begin{equation}
\label{field}
 {\vec{E}_3}({r},t)= \hat{e}_{3}\mathcal{E}_{3}(\vec{r})~e^{- i\left(\omega_3 t-k_3z \right)} + {c.c.}\,,
\end{equation}
where, $\mathcal{E}_{3}(\vec{r})$ is constant amplitude, $\omega_3$ is the frequency of the microwave field.
In the presence of three coherent fields, the  Hamiltonian of the system under the electric dipole and rotating-wave approximation can be expressed as,
\begin{subequations}
\label{Hschroed}
\begin{align}
H =& H_0 + H_I\,,\\
H_0 =& \hbar\omega_{43}\ket{4}\bra{4} - \hbar\omega_{23} \ket{2}\bra{2} - \hbar\omega_{13} \ket{1}\bra{1}\,,\\
H_I =& - ( \ket{4}\bra{1} \mf{d}_{41}\cdot\mathcal{E}_1e^{- i\left(\omega_1 t-  k_1 z\right )}  \nonumber \\
 &  + \ket{4}\bra{3} \mf{d}_{43}\cdot\mathcal{E}_2e^{- i\left(\omega_2 t-  k_2 z\right )}\nonumber\\   
 &  + \ket{3}\bra{2} \mf{d}_{32}\cdot\mathcal{E}_3e^{- i\left(\omega_3 t-  k_3 z\right )}\,+\,\text{H.c.})\,,
 \end{align}
\end{subequations}
The time dependent parts of the above Hamiltonian can be removed by use of unitary transformation,  
\begin{subequations}
\begin{align}
  W&=e^{-\frac{i}{\hbar}U t} \,,\\
  U&=\hbar \omega_2 \ket{4}\bra{4} - \hbar \omega_3 \ket{2}\bra{2} - \hbar (\omega_1-\omega_2)\ket{1}\bra{1}\,.
\end{align}
\end{subequations}
Now, we can rewrite transformed Hamiltonian as
\begin{align}
 {V}/{\hbar}  = & -\DC\ket{4}\bra{4} + \DM\ket{2}\bra{2} + (\DP-\DC)\ket{1}\bra{1}  \notag \\
 &   - (g\ket{4}\bra{1} + G \,\ket{4}\bra{3}   \,+ \Omega \,\ket{3}\bra{2} \notag \\ 
& \,+\,\text{H.c.})\,,
 \label{Heff}
\end{align}
where $\DP = \omega_1 - \omega_{41}$, $\DC = \omega_2 - \omega_{43}$, $\DM = \omega_2 - \omega_{32}$ are the single-photon detunings and 
\begin{equation}
\label{field}
g=\frac{\vec{d}_{41}\cdot\vec{\mathcal{E}}_{\rm{1}}e^{ik_1 z}}{\hbar},~~G=\frac{\vec{d}_{43}\cdot\vec{\mathcal{E}}_{\rm{2}}e^{ik_2 z}}{\hbar},~~ \Omega=\frac{\vec{d}_{32}\cdot\vec{\mathcal{E}}_{\rm{3}}e^{ik_3 z}}{\hbar}\nonumber
\end{equation}
are the Rabi frequencies of the probe, control and the microwave fields, respectively.
The atomic transition frequencies and the corresponding dipole moment matrix elements are denoted by $\omega_{ij}$ and ${\vec{d}}_{ij}$, respectively.
\subsection{Dynamical equations}
We use Liouville equation to incorporate  the coherent and incoherent processes of the atomic system.
Thus the dynamics of the system is governed by the  following Liouville equation
\begin{align}
\label{master}
\dot{\rho}=-\frac{\ii}{\hbar}\left[V,\rho\right]+\mathcal{L}\rho\,.
\end{align}
where the second term represents the incoherent processes that can be determined by
\begin{align}
\mathcal{L}\rho = &\mathcal{L}_{\gamma}\rho+\mathcal{L}_{d}\rho+\mathcal{L}_{r}\rho\,,
 \label{decay}
\end{align}
with
\begin{align}
\mathcal{L}_{\gamma}\rho = &-\sum\limits_{i=1}^3 \frac{\gamma_{i}}{2}\left(\ket{4}\bra{4}\rho-2\ket{i}\bra{i}\rho_{44}+\rho\ket{4}\bra{4}\right)  \,,\nonumber\\
 \mathcal{L}_{d}\rho = &-\sum\limits_{i=1}^3\sum\limits_{i\neq j=1}^3 \frac{\gamma_{c}}{2}\left(\ket{i}\bra{i}\rho-2\ket{j}\bra{j}\rho_{ii}+\rho\ket{i}\bra{i}\right)  \,,\nonumber\\
 \mathcal{L}_{r}\rho = &\mathcal{L}_{14}\rho  +  \mathcal{L}_{41}\rho \,,\nonumber\\
 \mathcal{L}_{14}\rho = &- \frac{r}{2}\left(\ket{4}\bra{4}\rho-2\ket{1}\bra{1}\rho_{44}+\rho\ket{4}\bra{4}\right) \,,\nonumber\\
 \mathcal{L}_{41}\rho = &- \frac{r}{2}\left(\ket{1}\bra{1}\rho-2\ket{4}\bra{4}\rho_{11}+\rho\ket{1}\bra{1}\right)\,.\nonumber
\label{idecay}
\end{align}
The first term of Eq.(\ref{decay}) refers to the radiative decay from excited state $|4\rangle$ to ground states $|j\rangle$ as labelled by $\gamma_{j}$. 
The second term, $\mathcal{L}_{d}\rho$, represents pure dephasing for the coherence $\rho_{ij}$ due to collision with rate $\gamma_{c}$. 
The incoherent pumping between levels $\ket{1}$ and $\ket{4}$ with rate $r$ is descryibed by $\mathcal{L}_{r}\rho$.
The dynamics of the population and atomic coherences in the four level system can be described by the following set of density matrix equations
\begin{subequations}
\label{Full_density}
\begin{align}
 \dot{\rho}_{11}=&-r\rho_{11}+r\rho_{44}+\gamma_{1}\rho_{44}+ \ii g^* \rho_{41}
 - \ii g \rho_{14} \,,\\
 \dot{\rho}_{22}=&  \gamma_{2}\rho_{44} + \ii \Omega^* \rho_{32}  - \ii \Omega
 \rho_{23}\,,\\
\dot{\rho}_{33}=&  \gamma_{3}\rho_{44} + \ii \Omega \rho_{23}  - \ii \Omega^*
 \rho_{32}+ \ii G^*\rho_{43} - \ii G\rho_{34}\,,\\
\dot{\rho}_{44}=&-\dot{\rho}_{11}-\dot{\rho}_{22}-\dot{\rho}_{33} \,,\\
 \dot{\rho}_{21}=&-\left[\frac{r}{2} + \gamma_{21} - \ii (\Delta_1-\Delta_2-\Delta_3)\right]\rho_{21} + \ii \Omega^*  \rho_{31} \nonumber\\
&- \ii g \rho_{24} \,,\\
 \dot{\rho}_{23}=&-\left[\gamma_{23} + \ii \Delta_3\right]\rho_{23}
 - \ii G\rho_{24} + \ii \Omega^*(\rho_{33} - \rho_{22})\,,\\
 \dot{\rho}_{24}=&-\left[ \gamma_{24} + \ii (\Delta_2+\Delta_3)\right]\rho_{24}
 - \ii g^* \rho_{21}- \ii G^* \rho_{23} \nonumber\\
& + \ii \Omega^*\rho_{34} \,,\\
 \dot{\rho}_{31}=&-\left[\frac{r}{2} + \gamma_{31} + \ii (\Delta_2-\Delta_1)\right]\rho_{31}
 + \ii \Omega\rho_{21}- \ii g^* \rho_{34} \nonumber\\
& + \ii G^*\rho_{41} \,,\\
  \dot{\rho}_{34}=&-\left[ \gamma_{34} - \ii \Delta_2\right]\rho_{34}
 - \ii g^*\rho_{31}+ \ii \Omega\rho_{24} \nonumber\\
& - \ii G^*(\rho_{33} - \rho_{44}) \,,\\
  \dot{\rho}_{41}=&-\left[\frac{r}{2} + \gamma_{41} - \ii \Delta_1\right]\rho_{41}
 + \ii G \rho_{31} - \ii g(\rho_{11} - \rho_{44}) \,,\\
\dot{\rho}_{ij}=&\dot{\rho}_{ji}^*\,.
 \end{align}
\end{subequations}
where, the overdots stand for time derivatives and $``*"$ denotes complex conjugate. 
The total dephasing rate of the atomic coherences is given by ${\gamma}_{ij} = {\gamma}_{c} +{{\gamma}_{i}}/{2}$.
\subsection{Perturbative analysis}
We adopt steady state solutions of the master equations (\ref{Full_density}) to study the response of the medium. 
The equations (\ref{Full_density}) can be solved to all orders in the control and probe field provided both the fields have approximately equal amplitude \cite{om}. 
However, in the spirit of weak probe field limit, we calculate the coherences and populations to the first order in $g$ and to all order in control field $G$ and 
microwave field $\Omega$. Hence the steady state solutions of the density matrix equations can be written in the form of
  \begin{equation}
  \rho_{_{ij}}=\rho_{_{ij}}^{(0)}+g\rho_{_{ij}}^{(+)}+g^*\rho_{_{ij}}^{(-)},
  \end{equation}
where, $\rho_{ij}^{(0)}$ describes the solution in the absence of the probe field. 
The second and third terms denote the solutions at positive and negative frequencies of the probe field. 
We now substitute the above expression in equations (\ref{Full_density}) and equate the coefficients of $g$, $g^{*}$ and the constant terms.
Thus, we obtain a set of sixteen coupled simultaneous equations.
The solutions of simultaneous equations which are relevant for susceptibility expression are given in Appendix~\ref{app-A}.
Now, the steady state value of the atomic coherence ${\rho}_{41}^{(+)}$ will yield susceptibility $\chi_{41}$ at frequency $\omega_1$
%
\begin{align}
\label{mollow_chi}
{\rho}_{41}^{(+)} &=i\left(\frac{{(\Gamma_{21}\Gamma_{31}+\Omega^2)({\rho}_{11}^{(0)}-{\rho}_{44}^{(0)})}+A G^2}{\Gamma_{41}(\Gamma_{21}\Gamma_{31}+\Omega^2)+\Gamma_{21}G^2}\right)\,, 
\end{align}
with
\begin{align}
{A} &= \frac{B({\rho}_{44}^{(0)} - {\rho}_{33}^{(0)})+C({\rho}_{33}^{(0)} - {\rho}_{22}^{(0)})}{(\Gamma_{23}(\Gamma_{24}\Gamma_{34}+\Omega^2)+\Gamma_{34}G^2)}\,,\nonumber \\
{B} &=(\Gamma_{21}(\Gamma_{23}\Gamma_{24}+G^2)-\Gamma_{23}\Omega^2)\,,\nonumber \\
{C} &= (\Gamma_{21}+\Gamma_{34})\Omega^2\,.\nonumber
\end{align}
where 
${\Gamma}_{21}=\left[r/2 + \gamma_{21} - \ii (\Delta_1-\Delta_2-\Delta_3)\right]$, ${\Gamma}_{23}=\left[\gamma_{23} + \ii \Delta_3\right]$, 
${\Gamma}_{24}=\left[ \gamma_{24} + \ii (\Delta_2+\Delta_3)\right]$, ${\Gamma}_{31}=\left[r/2 + \gamma_{31} + \ii (\Delta_2-\Delta_1)\right]$,
${\Gamma}_{34}=\left[ \gamma_{34} - \ii \Delta_2\right]$, and ${\Gamma}_{41}=\left[r/2 + \gamma_{41} - \ii \Delta_1\right]$.
 For the simplicity, we have assumed equal decay rates from excited state, $\gamma_{1}=\gamma_{2}=\gamma_{3}=\gamma$ and coherence dephasing rates $\gamma_{41}=\gamma_{24}=\gamma_{34}\approx\gamma$, $\gamma_{21}=\gamma_{31}=\gamma_{23}\approx\gamma_{c}=\Gamma$.
We now express the macroscopic polarization of the medium in terms of both the atomic coherences as well as the susceptibility as
\begin{align}
\label{polarization}
\vec{\mathcal{P}}_1&=\mathcal{N}\left(\vec{d}_{41}\rho_{41}^{(+)}e^{-\ii\omega_1 t}+c.c.\right)\notag\\
&=\left(\chi_{41} \hat{e}_{1}\mathcal{E}_1e^{-\ii\omega_1 t}+c.c.\right)\,,
\end{align} 
where $\mathcal{N}$ is the density of the atomic medium. 
 Now Eq.~(\ref{mollow_chi}) and (\ref{polarization}), will yield the linear response of the medium as
\begin{align}
\label{chi_41}
{\chi}_{41}(\DP)&=\frac{\mathcal{N}|d_{41}|^2}{{\hbar}}{\rho}_{41}^{(+)} .
\end{align}
The real and imaginary parts of the susceptibility ${\chi}_{41}$ in Eq.~(\ref{chi_41}) gives the dispersion and absorption of the medium respectively. 
The optical properties of the medium can be manipulated coherently by proper consideration of spatial shape and intensity of the different applied fields. 
The effect of different fields such as optical, microwave and incoherent pump field on the medium properties are in sequence in the results and discussions section.
\subsection{Beam propagation equation with paraxial approximation}
The spatial dynamics of the probe and control fields along the $z$-direction of the medium
is governed by the Maxwell's wave equations.
The wave equation under slowly varying envelope and paraxial wave approximations can result the beam propagation equation. 
The  spatial evolution equations for the probe and control fields are obtained as  
\begin{subequations}
\begin{align}
 \frac{\partial g}{\partial z}
   &= \frac{\ii }{2{k_1}} \left( \frac{\partial^2 }{\partial x^2}
      + \frac{\partial^2 }{\partial y^2} \right) g + 2i{\pi}k_1{\chi}_{41}\,{g} \,,\label{probe} \\
 \frac{\partial G}{\partial z}
   &= \frac{i}{2{k_2}} \left( \frac{\partial^2 }{\partial x^2}
      + \frac{\partial^2 }{\partial y^2} \right) G \,\label{control}.
\end{align}
\end{subequations}
The terms within the parentheses on the right hand side of Eq.~(\ref{probe}) and  Eq.~(\ref{control}) are related with  transverse variation of the laser beam. 
These terms account for the diffraction either in free space or in the medium. 
The second term on the right hand side of Eq.~(\ref{probe}) is responsible for the dispersion and absorption or gain of the probe beam. 
Note that the effects of the atomic coherences on the control beam propagation are very negligible  under the weak probe field \cite{om}. 
Therefore, we study the effect of both diffraction and dispersion for the spatial evolution of the probe beam  where  we include only the effect of diffraction for the control beam dynamics. 
\section{Results and Discussions }
\subsection{Susceptibility with continuous wave fields}
\begin{figure}[t]
\includegraphics[width=1.0\columnwidth]{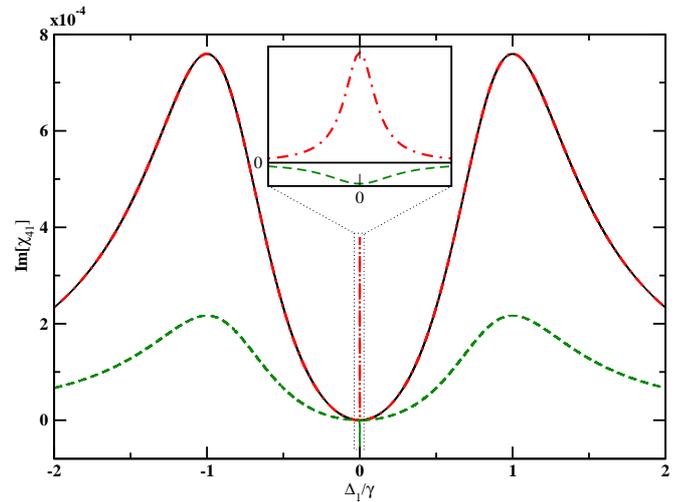}
\caption{\label{fig:Fig2} (Color online)
The variations of the imaginary part of the probe susceptibility with the detuning $\DP$ in the presence and absence of both microwave field and incoherent pump is plotted. 
The zoomed part of the absorption spectrum corresponds to medium loss, gain or transparency at the line center  is shown in the inset. 
The corresponding parameters for these regimes are: ${\Omega}=0.01{\gamma}$, $r=0$ (red dot-dashed line), ${\Omega}=0.01{\gamma}$, $r=0.0005\gamma$ (green dashed line), and ${\Omega}=0{\gamma}$, $r=0$ (black solid line). The common parameters are $G=1.0{\gamma}$, $\DC=\DM=0$, $\Gamma=0.0001\gamma$, $\gamma=3\pi\times10^6$ rad/sec, and $N=5{\times}10^{11}$ atoms/cm$^{3}$.}
\end{figure}
We first study the atomic coherences by using cw optical and microwave fields at steady-state condition.
The quantum interference of atomic coherences induces  EIT,  MIA and LWI in our system. 
The characteristic of these quantum interference phenomena  is illustrated in Fig.~\ref{fig:Fig2}.
In Fig.~\ref{fig:Fig2} we have plotted the variations of the imaginary part of the probe susceptibility with probe field detuning $\DP$ in the presence and absence of both microwave and incoherent pump fields.  
In the absence of both microwave and incoherent pump fields  four-level system reduces to three-level $\Lambda$ system with a weak probe and a strong control field. 
The probability amplitudes of two arms of the $\Lambda$ system leads to destructive interference. 
This interference enable us to cancellation of absorption of probe field provided two-photon resonance condition is fulfilled  as shown in Fig.~\ref{fig:Fig2}. 
This phenomenon  is known as  EIT. 
In EIT, a single transparency window is accompanied by two absorptive peaks which originates from the strong control field. 
Now  this single transparency window can be split into double transparency windows by the use of the microwave field. 
It is clear from Fig.~\ref{fig:Fig2} that the double transparency window is accompanied with very narrow absorption peak.
This peak occurs due to the double dark states formed by microwave field at three-photon resonance condition. 
Furthermore, the position and width of these two transparency windows strongly depend  on the intensity of microwave field. 
Now a relatively weak incoherent pump acting along the probe transition can switch the absorption peak to the gain dip.
The second term in the numerator of Eq.~(\ref{mollow_chi}) is responsible for gain around line center. 
This gain characteristic is illustrated by green dashed line line in Fig.~\ref{fig:Fig2}. 
At three photon resonance the second term is negative and is lager than the first term which changes the properties of the medium from absorption into gain.
Thus the presence of both weak microwave and incoherent pump fields is able to produce a gain window for the medium. 

\begin{figure}[t]
\includegraphics[width=1.0\columnwidth]{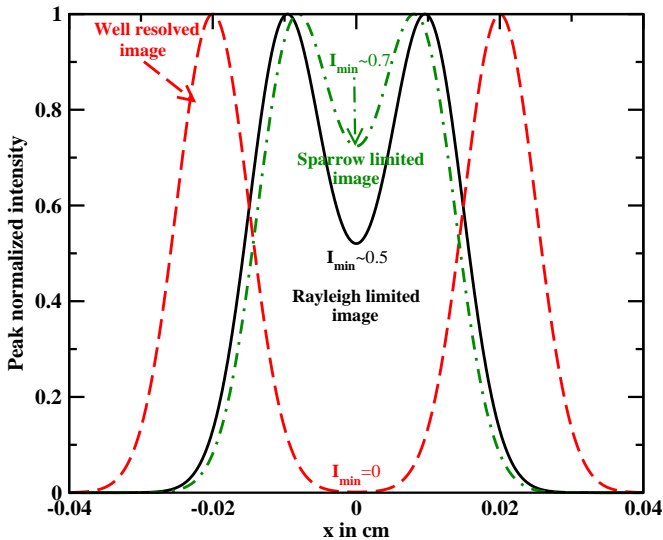}
\caption{ \label{fig:Fig3} (Color online) 
Spatial intensity variation of the control image is plotted against the transverse axis $x$ with $y=0$ at entry face of the vapor cell. 
The Rayleigh limited and Sparrow limited control image are formed by choosing $a_1=-a_2=0.01$cm and $a_1=-a_2=0.009$cm, respectively . 
The individual peaks can be well resolved by changing $a_1=-a_2=0.02$cm.
The common parameters of two graphs are $G_{0}=1\gamma$, and ${w_c}=100{\mu}$m.}
\end{figure}
\begin{figure}[b]
\includegraphics[width=1.0\columnwidth]{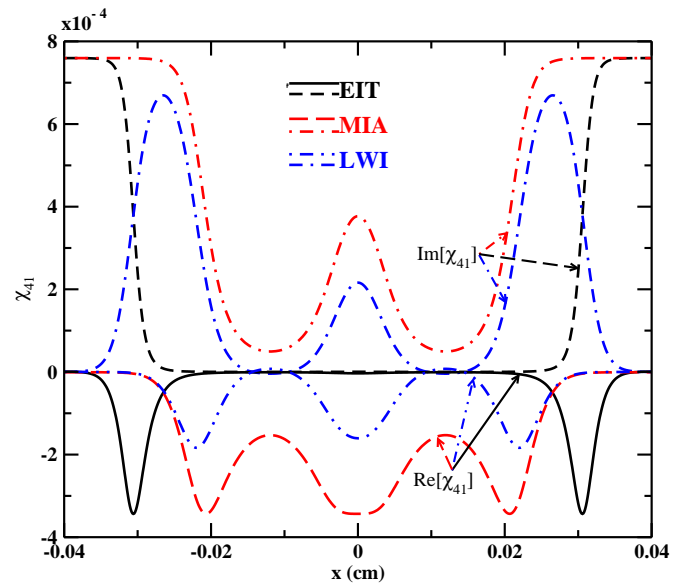}
\caption{\label{fig:Fig4} (Color online) The Spatial variation of the real (Re$[\chi_{41}]$) and imaginary (Im$[\chi_{41}]$) parts of $\chi_{41}$. 
The plots are shown against the transverse axis coordinate $x$ of the control beam for $y=0$ plane.  
The different curves are for three different set of parameters: ${\Omega}=0.015{\gamma}$, $r=0$, $\DP=0.001{\gamma}$ (red long dashed, and dot-dashed lines); ${\Omega}=0.015{\gamma}$, $r=0.0005\gamma$, $\DP=0.001{\gamma}$ (blue dashed double-dot, and dot double-dashed lines), and ${\Omega}=0$, $r=0$,  $\DP=-0.001{\gamma}$ (black solid, and short dashed lines). 
The control beam parameters are $G_{0}=1\gamma$, ${w_c}=100{\mu}$m, and $a_1=-a_2=0.012$cm.}
\end{figure}
\subsection{Susceptibility with inhomogeneous control field}
In this section, we discuss the effect of spatial inhomogeneous field on linear susceptibility given in Eq.~(\ref{chi_41}).
For this purpose, we change the control field profile from cw to spatially inhomogeneous field while keeping rest of the fields as cw for further study. 
The spatially inhomogeneous transverse profile of the control field is a combination of more than one Gaussian peak. 
At $z=0$, the control beam can be written as,
\begin{align}
\label{shape_drive1}
G(x,y)=&{G}_0\sum\limits_{i=1}^n \:e^{-\frac{\left [(x-a_i)^2+y^2\right]}{w_c^{2}}}\,,
\end{align}
where, $G_0$ is initial peak amplitude, $w_c$ is beam width and $a_i$ are the individual peak position. 
The full-width-at-half-maximum (FWHM) of individual peak is $\sqrt{2ln2}w_c$. 
Figure {\ref{fig:Fig3}} shows the intensity distribution of the control field
against radial position $x$ at the entry face of the medium. 
The overlapping of two peaks gives rise to a central minimum with non-zero intensity as shown in Fig.\ref{fig:Fig3}. 
The Rayleigh-limited  or Sparrow limited control images can be formed when the intensity of the peak normalized central minimum  is $I_{min}\sim0.5$ or $\sim 0.7$, respectively.
The resolution of the diffraction limited images can be improved by reducing the central minimum intensity to zero.
Thus, by increasing the peak separation or by decreasing the width of the individual peak can able to create high resolution image.

The spatially modulated control field perturbs the probe beam susceptibility along the transverse direction as shown in Fig.~\ref{fig:Fig4}.
Fig.~\ref{fig:Fig4} illustrates the spatial variation of the real and imaginary parts of $\chi_{41}$ as a function of the transverse axis $x$ for $y=0$ plane.
The very special inhomogeneous character of dispersion Re$[\chi_{41}]$ and absorption Im$[\chi_{41}]$ causes the spatial modulation in phase and amplitude for the probe field, respectively.
Since the phase of probe beam is influenced by the co-propagating control beam, therefore, this phase modulation is termed as cross phase modulation (XPM) \cite{agrawal}. 
The mutual coupling between the optical beams is attributed to XPM which causes focusing to the probe beam.
The amplitude modulation results in attenuation or gain to the probe beam.

The curves of Fig.~\ref{fig:Fig4} represent three different cases of EIT, MIA, and LWI, respectively. 
It is clear from Fig.~\ref{fig:Fig4} that for MIA and LWI cases two transparency windows are formed at higher intensity regions 
whereas absorption occurs in relatively low intensity regions of control field $G$ defined by two Gaussian modes using Eq.~(\ref{shape_drive1}).
The real part of the susceptibility is maximized at these higher intensity regions. 
This resembles two parallel waveguide like structures  with claddings (0.0075 cm$ \gtrsim |x| \gtrsim $ 0.0175 cm) and cores (0.0175 cm$ \gtrsim |x| \gtrsim $ 0.0075 cm). 
In order to have a perfect wave-guiding, there should be a high contrast between core and cladding. 
In case of EIT,  it is evident from Fig.~\ref{fig:Fig4} that a single transparency window is formed and the variation in refractive index around $x=0$ is very small. 
Therefore, the single transparency window is failed to create two parallel waveguide.
As a result, EIT is not suitable to separate out the modes with high resolution. 
However, in the case of MIA, one can see a sharp variation in refractive index (red long dashed line) around $x=0$, with a rapid increase in contrast from core to cladding. 
But there is reasonable increase in absorption in the region between 0.0175 cm$ \gtrsim |x| \gtrsim $ 0.0075 cm of the doublet compared to EIT. 
This increment will reduce transmission of the probe beam and therefore, its visibility seems to be restricted.
\begin{figure}[t]
\subfigure[]
 {
 \centering
   \includegraphics[width=1\columnwidth]{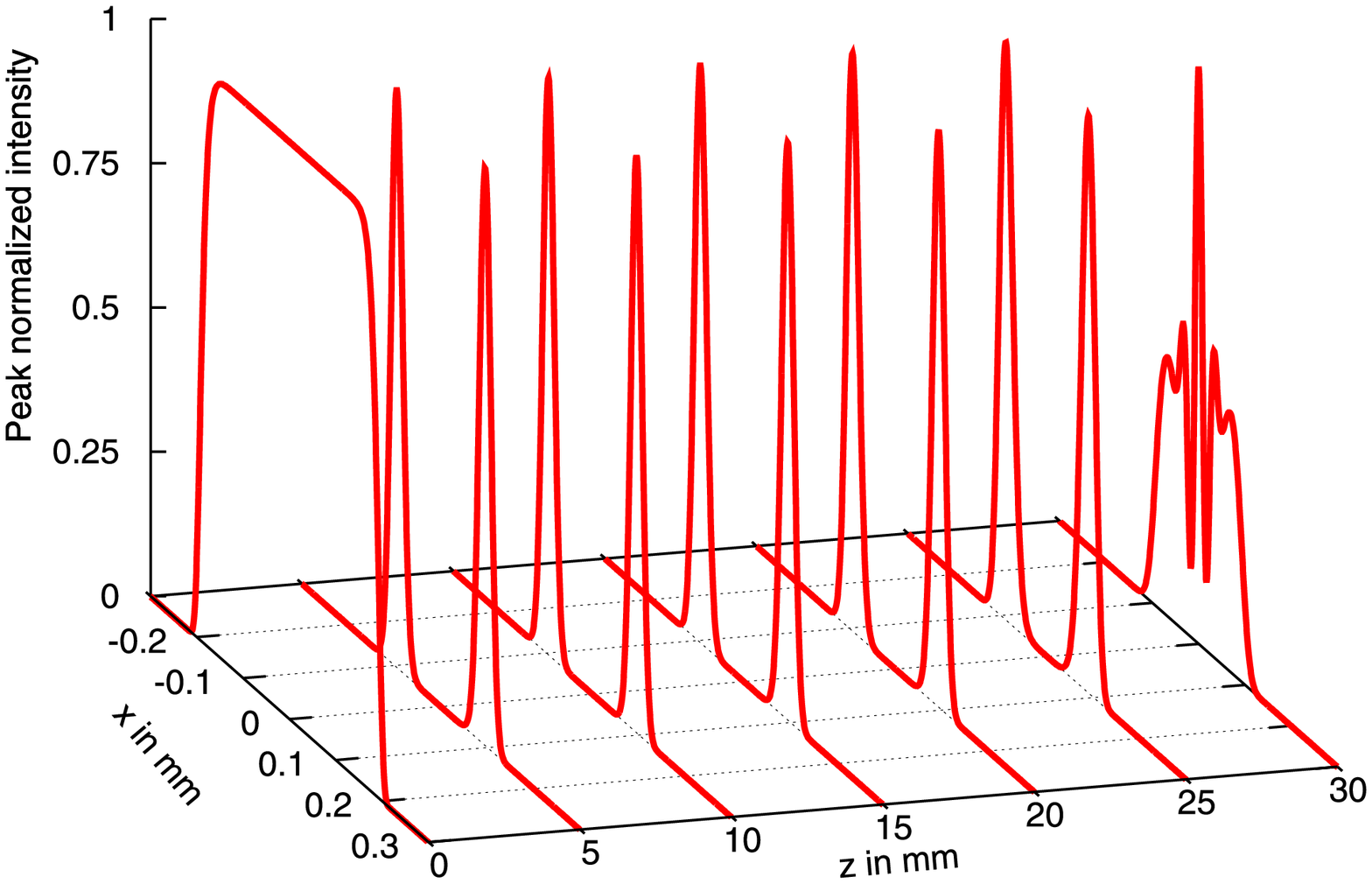}
   \label{fig:Fig5a}
 }
\subfigure[]
 {
 \centering
 \includegraphics[width=1\columnwidth]{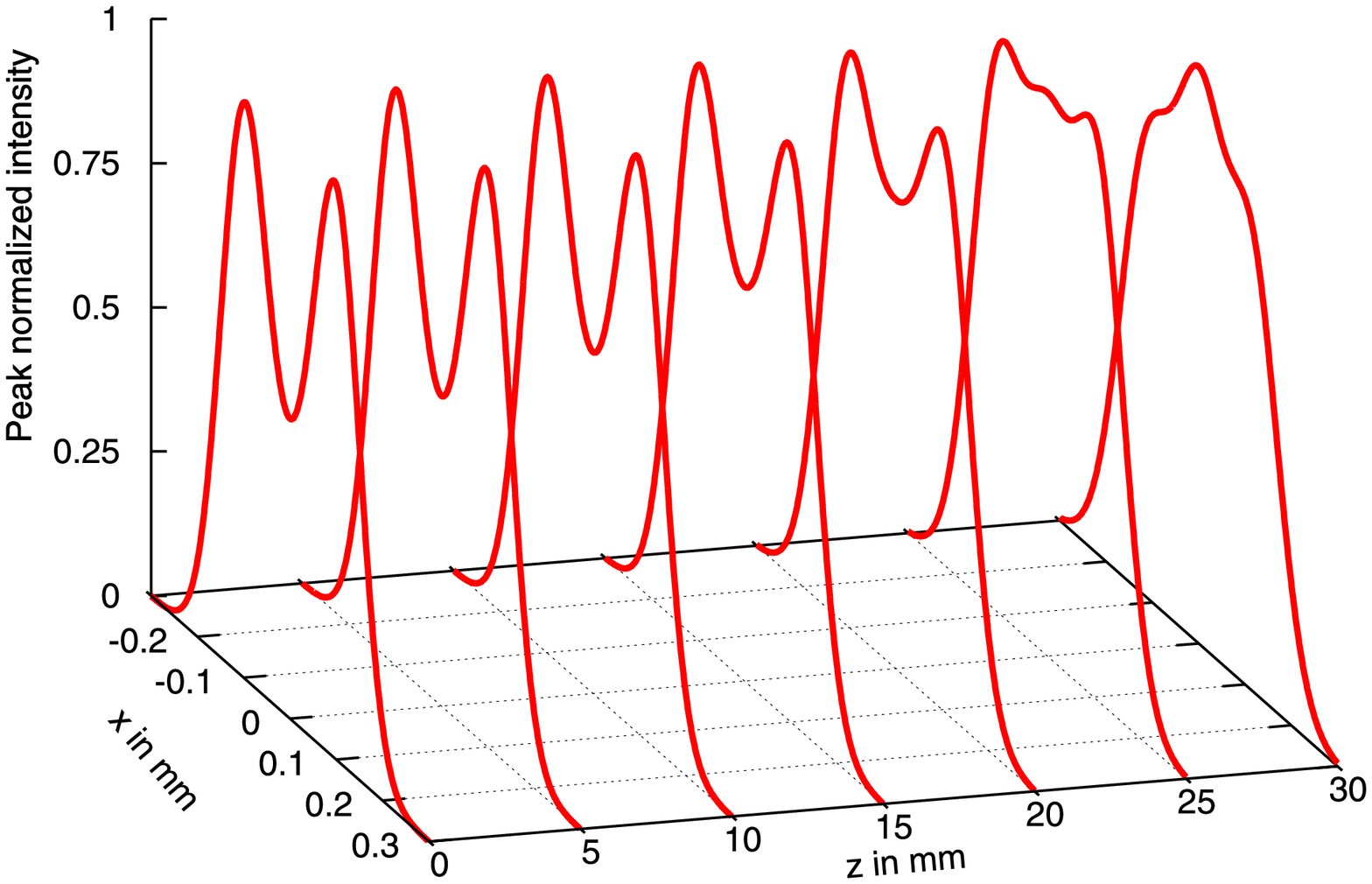}
   \label{fig:Fig5b}
 }
\caption{\label{fig:Fig5}(Color online)
In panel (a), the spatial evolution of probe beam profile is shown against the transverse coordinate $x$ for $y=0$ plane at different propagation distances $z$. 
In panel (b),  the peak-normalized intensity profile of the control beam is shown at different propagation distances z. 
The parameters are chosen as follows: ${\Omega}=0.018{\gamma}$, $r=0.00075{\gamma}$, $\DP=0.001{\gamma}$. The control beam (Rayleigh limited) parameters are same as in Fig.(\ref{fig:Fig3}).}
\end{figure}
\begin{figure}[b]
\includegraphics[width=1.0\columnwidth]{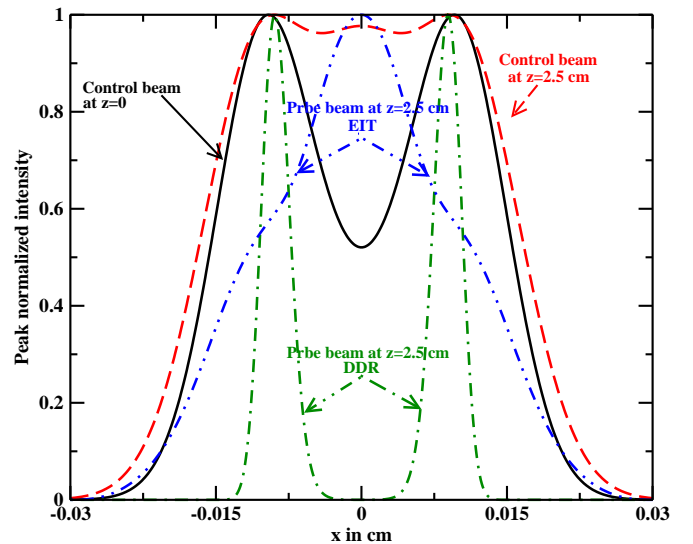}
\caption{ \label{fig:Fig6} (Color online) A comparison study of EIT and DDR with incoherent pump for cloning of the just resolved control images onto the probe beam 
at the output of the vapor cell with length $L=2.5$ cm. The parameters are same as in Fig.(\ref{fig:Fig4})}
\end{figure}

Interestingly, in case of LWI, the refractive index contrast between core and cladding is higher than the other two cases. 
This contrast enhancement causes  strong focusing of the probe beam towards the center of the two peaks of the control field.  
As a result the width of the probe beam becomes narrow which can improve  the contrast of the cloned image on the probe field.  
Also the two deeps of the doublet changes from absorption into gain can produce the enhancement  of the cloned beam transmission. 
Hence the weak probe beam is not only guided or focused but also amplified in order to preserve the information during the propagation through the optical medium.
This is the key mechanism of cloning the un-resolvable or just-resolvable control field profile to the probe field with high resolution.
In the following, we use the inhomogeneous susceptibility for LWI case to illustrate the improvement of the resolution of the cloned of images of the control field onto the probe field. 
\subsection{Beam propagation dynamics}
We numerically integrate the paraxial wave equations (\ref{probe}) and (\ref{control}) by using a higher order split operator method~\cite{Shen} to study the propagation dynamics of both control and probe beams. 
First we explore the cloning of Rayleigh limited control beam onto the probe beam in presence of both microwave and incoherent pump fields. 
For this purpose, we set  ${w_c}=100{\mu}$m and $a_1=-a_2=0.01$cm in the Eq.(\ref{shape_drive1}).
The results for the spatial evolution of the control and the probe profiles throughout the medium are shown in Fig.~\ref{fig:Fig5}.
It is clear from Fig.~\ref{fig:Fig5a}  that within a very short distance, the control field structure is mapped on to  the probe with central minimum reduced to zero. 
As a result, the finesse, which is the ratio of the spacing between peaks to the width of peaks of the transmitted probe beam at $z=2.5$ cm, 
is 4 times smaller than initial control beam finesse. 
We also find that the integrated transmission of the output probe beam at $z=2.5$ cm is about $98\%$. 
The probe beam transmission can be changed by changing the incoherent pump field rate $r$.
Figure~\ref{fig:Fig5b}  depicts the intensity profile of the control beam at different propagation distances $z$.
We find that the the shape of the control beam is gradually distorted as it propagates through the medium due to diffraction. 
As a consequence, control beam induced waveguide structure in the medium is modified.
Accordingly the shape of the cloned beam starts experiencing diffraction after $z=2.5$ cm propagation distance as shown in Fig.\ref{fig:Fig5a}. 
Long distance diffractionless cloned image propagation can be achieved by considering tightly focused control beam \cite{howell_09} or self-reconstructing Bessel control beam\cite{Fahrbach}.
\begin{figure}[t]
\centering
\subfigure[]
 {
   \includegraphics[width=0.7\columnwidth,angle=270]{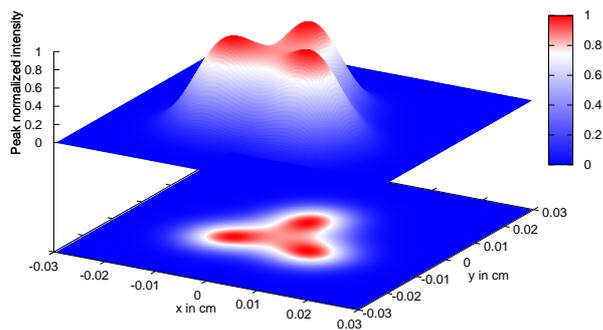}
   \label{fig:Fig7a}
 }
\subfigure[]
 {
   \includegraphics[width=0.7\columnwidth,angle=270]{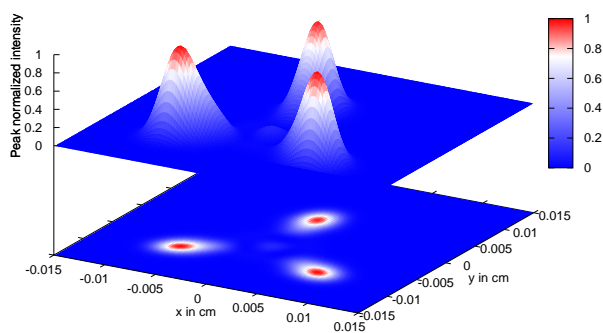}
   \label{fig:Fig7b}
 }
\caption{\label{fig:Fig7}(Color online)Picture (a) shows 3-D  intensity profile of the input control beam. 
Picture (b) shows the transmitted probe beam at the output of a 1~cm long medium. 
The parameters are as in Fig.~\ref{fig:Fig5} except location of the three peaks are $(-0.009, -0.009)$, $(0.009, -0.009)$, $(0.0, 0.0066)$cm, and $\Omega=0.02\gamma$, $r=0.00073\gamma$.} 
\end{figure}
\begin{figure}[b]
\centering
\subfigure[]
 {
   \includegraphics[width=1\columnwidth,angle=0]{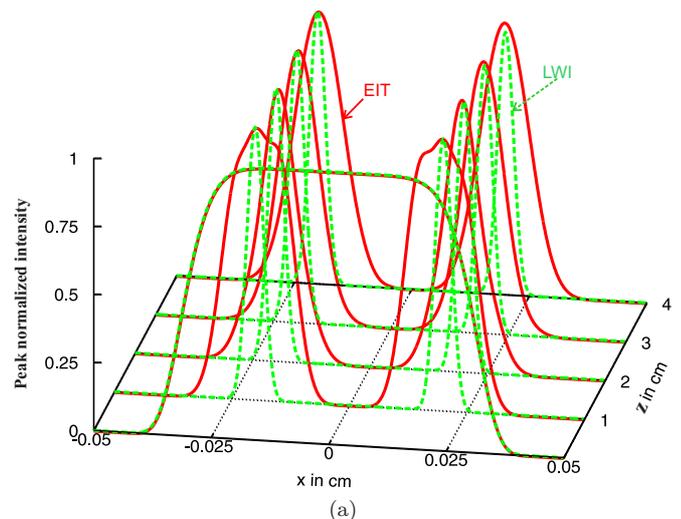}
   \label{fig:Fig8a}
 }
\subfigure[]
 {
   \includegraphics[width=1\columnwidth,angle=0]{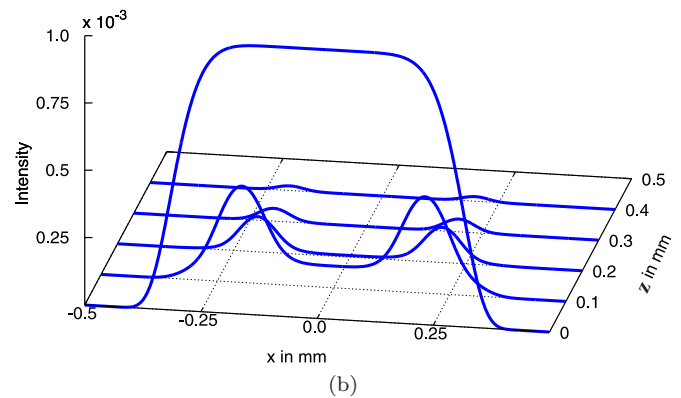}
   \label{fig:Fig8b}
 }
\caption{\label{fig:Fig8}(Color online) The intensity profile of the probe field transmission is shown against transverse axis $x$ with $y=0$ at different propagation distances $z$. 
Top panel shows probe beam is turned on in both EIT and LWI cases. 
The lower panel shows probe beam is turned off in MIA situation. 
The initial profile of the control field contains two well resolved Gaussian peaks with location $a_1=-a_2=0.02$cm as in Eq.~(\ref{shape_drive1}). 
The parameters are used in the different phenomena for spatial optical switching as follows: in EIT case ($\Omega=0, r=0, \Delta_1=-0.005\gamma$)
in LWI case ($\Omega=0.01\gamma, r=0.0001\gamma, \Delta_1=0.0001\gamma$), and in MIA case ($\Omega=0.01\gamma, r=0, \Delta_1=0.0001\gamma$). }
\end{figure}

Figure \ref{fig:Fig6} compares the cloning mechanism in presence and absence of both microwave and incoherent pump fields.
The Rayleigh limited control field structure generated a double transparency window and a single transparency window for DDR and EIT system, respectively.
It is clear from Fig. \ref{fig:Fig6} that the double transparency window enable to perfectly clone the control image whereas  single transparency window failed to clone the control image to the transmitted probe beam.
We also notice that the DDR induced waveguide structure can support the propagation of cloned probe beam without any diffraction. 
In contrast, for EIT case, the transmitted probe beam suffers severe distortion due to lack of parallel waveguide like structure inside the medium. 
Hence EIT based mechanism has limitation to clone unresolved or just resolved control image onto probe beam without loss of generality.

Next, we demonstrate how the microwave and incoherent pump fields offer the unprecedented control over the image cloning for unresolved images.
For this purpose, we consider more complex structure of control beam consisting of three Gaussian peaks. 
Fig. \ref{fig:Fig7} shows the radial distribution of the input Sparrow limited control beam (at z = 0) and output probe beam at z = 1 cm. 
As in Fig. \ref{fig:Fig7b} it can be seen that  the cloned probe images contains three distinguishable peaks even though the control beam profile is unresolved. 
Surprisingly,  the integrated transmission intensity of the cloned probe image is approximately 74\%. 
Thus microwave and incoherent pump fields allow one to cloned the diffraction limited control field image onto the probe beam with improved spatial resolution and high transmission.
We also verified that the resolution enhancement of cloned images can be possible even for Rayleigh limited control images with the Bessel as well as non-Gaussian shape.
These studies may be useful for practical applicability such as  optical microscope, quantum metrology and quantum imaging\cite{application}. 
\subsection{Spatial optical switchting}
Here, we show how the propagation dynamics of the probe beam can be controlled by switching the microwave field on and off. 
The well resolved control beam image is being considered for this demonstration.
The individual peak has width 100$\mu$m correspond to 4 cm Rayleigh length.
The spatially dependent control field assisted atomic waveguide can protect the feature of the cloned beam in a 4 cm long medium. 
Fig.\ref{fig:Fig8a} illustrates that the nondiffracting cloned probe beam propagation is possible inside the medium in both EIT as well as LWI system. 
We found that the width and the transmission of the cloned beam at $z=3$ cm are 25$\mu$m (100 $\mu$m) and 60\% (5\%) for LWI (EIT) mechanism. 
Therefore, the precise control of finesse and the contrast of the output cloned probe beam can be achieved by application of coherent fields and incoherent pump field interacting in a four level atomic medium.
Fig.\ref{fig:Fig8b} shows how the microwave induced absorption can be utilized to attenuate the probe beam gradually inside the medium in the absence of incoherent pump field.
Thus, microwave field which connects the lower level metastable states of four level system can switch off the probe beam propagation inside the medium.
This investigation can be applicable for all optical switching and logic gates \cite{hong,nie}. 
\section{\label{Conclusion}Conclusion}
In conclusion, we have revealed  a scheme to improve the resolution of the cloned image based on the quantum interference effects induced by interacting dark resonances. 
For this purpose, we have used four levels atomic system interacting with three coherent fields and an incoherent pump field. 
An atomic wave-guide structure is formed inside the medium by using a spatially modulated control field. 
The refractive-index contrast between core and cladding of the atomic waveguide can be increased by use of sharp absorption peak associated with double dark resonances. 
The high contrast  atomic wave-guide enables us to imprint the Rayleigh or Sparrow limited control images to probe field with high resolution. 
The transverse feature of control image is efficiently cast on to the probe field even though the control image suffers distortion due to the diffraction during the propagation. 
Our numerical result show that the propagation of high resolution cloned image is possible until the feature of the control image lost completely. 
We use incoherent pump field in order to increase the transmission of the cloned probe image. 
Finally, we have also demonstrated that spatial optical switching is possible by use of EIT, LWI and MIA mechanism.

\begin{acknowledgements}
{One of the authors (T.N.D) gratefully acknowledge funding by the Science and Engineering Board(SR/S2/LOP-0033/2010).}

\end{acknowledgements}

\appendix
\section{\label{app-A}Coefficients for susceptibility}

\begin{widetext}
\begin{align}
{\rho}_{11}^{(0)}=&\frac{(2(r+\gamma)|G|^2\Omega^2({\gamma}({\Gamma}{\gamma}+|G|^2)+{\Gamma}(({\DC}+{\DM})^2+{\Omega}^2)))}{D}\\
{\rho}_{22}^{(0)}=&(r(|G|^6+G^4(2\Gamma\gamma^2-2\gamma\DC\DM-2\gamma\DM^3+\Gamma\Omega^2-\gamma\Omega^2)+2\Gamma\Omega^2(\DC^4+2\DC^3\DM+\gamma^2\DC^3+\DC^2(2\gamma^2+\DM^2-2\Omega^2)\notag\\
&+2\DC\DM(\gamma^2-\Omega^2)+(\gamma^2+\Omega^2)^2)+|G|^2(\gamma\DM^4+(\gamma^2+\Omega^2)({\Gamma}^2\gamma-(\Gamma-2\gamma)\Omega^2)+\DM^2(\gamma(\Gamma^2+\gamma^2)\notag\\
&+2(\Gamma+2{\gamma})\Omega^2)+\DC^2({\Gamma}^2\gamma+\gamma\DM^2+(5\Gamma+2\gamma)\Omega^2)+\DC\DM(2{\Gamma}^2\gamma+2\gamma\DM^2+(7\Gamma+6\gamma)\Omega^2))))/{D}\\
{\rho}_{33}^{(0)}=&(r\Omega^2(2{\Gamma}\DC^4+4\Gamma\DC^3\DM+\DM^2(2\Gamma{\gamma}^2+(2\Gamma+\gamma)|G|^2)+\DC^2(2\Gamma\DM^2+(3\Gamma+2\gamma)G^2+4\Gamma(\gamma^2-\Omega^2))\notag\\
&+\DC\DM((5\Gamma+2\gamma)|G|^2+4\Gamma(\gamma^2-\Omega^2))+(|G|^2+2({\gamma}^2+\Omega^2))(\gamma|G|^2+\Gamma({\gamma}^2+\Omega^2))))/{D}\\
{\rho}_{44}^{(0)}=&\frac{(2r|G|^2\Omega^2({\gamma}({\Gamma}{\gamma}+|G|^2)+{\Gamma}(({\DC}+{\DM})^2+{\Omega}^2)))}{D}\\
D=&r\gamma |G|^6 +|G|^4\left(2r \gamma(\Gamma \gamma- \DM(\DC+\DM))+(2 \gamma^2+r(\Gamma +4 \gamma))\Omega^2\right)\notag\\
&+4r\Gamma \Omega^2 \left((\gamma^2+\DC^2)(\gamma^2 +(\DC+\DM)^2)+2(\gamma^2 -\DC(\DC+\DM))\Omega^2+\Omega^4\right)\notag\\
&+|G|^2 \left(r\gamma(\Gamma^2+\DM^2\right)\left(\gamma^2+(\DC+\DM)^2\right)\notag\\
&+(\gamma(2\Gamma\gamma^2 +r(\Gamma+2\gamma)^2)+2 \left(\Gamma\gamma+2r(3\Gamma+\gamma)) \DC^2+4(5r \Gamma+2r \gamma + \Gamma \gamma \right)\DC \DM \notag\\
&+(8r\Gamma+5r \gamma +2\Gamma \gamma) \DM^2 )\Omega^2+2 (\Gamma\gamma +2r( \gamma +\Gamma)\Omega^4))\,.
\end{align}
\end{widetext} 

\end{document}